\documentclass[a4paper,12pt,reqno]{amsart}

\usepackage{graphicx,amssymb,datetime,float,MnSymbol,currfile,tikz,multirow}
\usepackage[square,comma,numbers]{natbib}
\usepackage[foot]{amsaddr}
\usetikzlibrary{arrows}
\usetikzlibrary{decorations.markings}

\def\ra{\rightarrow}

\def\la{\leftarrow}

\def\aa{\leftrightarrow}

\def\ao{\leftarrow\!\!\!\!\!\multimap}

\def\oa{\mathrel{\reflectbox{\ensuremath{\ao}}}}

\newcommand{\comments}[1]{}

\def\nx{\overline{x}}
\def\nz{\overline{z}}

\tikzset{tt/.style={decoration={
  markings,
  mark=at position .485 with {\arrow{>}},
  mark=at position .515 with {\arrow{<}}},postaction={decorate}}}

\begin{document}

\title[]{Alternative Measures of Direct and Indirect Effects}

\author{Jose M. Pe\~{n}a$^1$}
\address{$^1$Link\"oping University, Sweden.}
\email{jose.m.pena@liu.se}


\maketitle

\begin{abstract}
There are a number of measures of direct and indirect effects in the literature. They are suitable in some cases and unsuitable in others. We describe a case where the existing measures are unsuitable and propose new suitable ones. We also show that the new measures can partially handle unmeasured treatment-outcome confounding, and bound long-term effects by combining experimental and observational data.
\end{abstract}

\section{Introduction}

Consider the following causal graph studied in \cite{Pearl2001,Pearl2009}, where $X$, $Z$ and $Y$ represents an applicant's gender, qualifications and hiring, respectively.
\begin{center}
\begin{tikzpicture}[inner sep=1mm]
\node at (0,0) (E) {$X$};
\node at (2,0) (D) {$Y$};
\node at (1,0) (Z) {$Z$};
\path[->] (E) edge (Z);
\path[->] (Z) edge (D);
\path[->] (E) edge[bend left] (D);
\end{tikzpicture}
\end{center}
The edge $X \ra Y$ represents that the hirer questions applicants about their gender, and that their answers may have an effect on hiring them. \citeauthor{Pearl2001} imagines a policy maker who may be interested in predicting the gender mix in the work force, if it were illegal for the hirer to question applicants about their gender. This quantity corresponds to the effect of gender on hiring mediated by qualifications. \citeauthor{Pearl2001} argues that the answer to this question lies in deactivating the direct path $X \ra Y$. He also argues that the answer can be realized by computing the average natural (or pure) indirect effect:
\[
NIE(X,Y)=E[Y_{\nx,Z_x}]-E[Y_{\nx}]
\] 
which is the difference between the expected outcomes under no exposure when the mediator takes the value it would under exposure and non-exposure, respectively. We agree with the answer to the question (i.e., deactivating $X \ra Y$) but not with its realization (i.e., deactivating $X \ra Y$ by computing $NIE(X,Y)$), because the reference value $\nx$ affects the outcome in this realization of the answer. This is problematic because it means that the direct path $X \ra Y$ is not really deactivated and, moreover, the answer depends on the level chosen as reference. In other words, this realization of the answer does not really correspond to the no-questioning policy being evaluated. The problems discussed here are shared by other classical measures of indirect effect such as the average total and controlled indirect effects, as well as by the interventional indirect effect measure proposed by \citet{Geneletti2007}: 
\[
IIE(X,Y)=E[Y_{\nx,\mathcal{Z}_x}]-E[Y_{\nx,\mathcal{Z}_{\nx}}]
\]
which compares the expected outcome under no exposure when $Z$ is drawn from the distributions $\mathcal{Z}_x$ and $\mathcal{Z}_{\nx}$ of $Z_x$ and $Z_{\nx}$.\footnote{Although $NIE(X,Y)$ and $IIE(X,Y)$ do not coincide in general, they coincide for the causal graph under study \cite{VanderWeeleetal.2014}.} However, this does not mean that these measures should be abandoned. Quite the opposite. They are informative when the reference value is clear from the context. For instance, if women suspect being discriminated by the hirer, then they may want to know if the probability of a woman getting hired would remain unchanged had she a man's qualifications. This is measured by $NIE(X,Y)$ with reference value $\nx$ set to ``woman". In summary, the existing measures of indirect effect are suitable in some cases and unsuitable in others. In this paper, we propose a new measure that does not require selecting a reference value.

More recently, \cite{Fulcheretal.2020} have introduced the population intervention indirect effect to measure the indirect effect of $X$ on $Y$ through the mediator $Z$:
\[
PIIE(\nx) = E[Y_{X,Z_X}] - E[Y_{X,Z_{\nx}}]
\]
which is the difference between the expected outcomes when the exposure and mediator take natural (observed) values and when the exposure takes natural value but the mediator takes the value it would under no exposure. Therefore, this measure is suitable when the exposure is harmful (e.g., smoking) and, thus, one may be more interested in elucidating the effect (e.g., disease prevalence) of eliminating the exposure rather than in contrasting the effects of exposure and non-exposure. The latter is considered irrelevant, because it is inconceivable that everyone will be exposed. In this paper, though, we are interested in the latter because it may be informative even when the interventions are inconceivable. For instance, the two interventions being contrasted in the gender discrimination example above (everyone is male and everyone is female) are both inconceivable, but their contrast is instrumental to decide whether the no-questioning policy should be introduced or not, as argued in \cite{Pearl2001,Pearl2009} (see also the previous paragraph).

The rest of the paper is organized as follows. We present our new measure of indirect effect in Section \ref{sec:measures}. We also present a new measure of direct effect. We illustrate them with an example. Moreover, we show that they can partially handle unmeasured treatment-outcome confounding, and bound long-term effects by combining experimental and observational data. Finally, we close with some discussion in Section \ref{sec:discussion}.

\section{Alternative Measures}\label{sec:measures}

Consider again the causal graph below, where $X$, $Z$ and $Y$ represents an applicant's gender, qualifications and hiring, respectively.
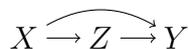
\begin{center}
\begin{tikzpicture}[inner sep=1mm]
\node at (0,0) (E) {$X$};
\node at (2,0) (D) {$Y$};
\node at (1,0) (Z) {$Z$};
\path[->] (E) edge (Z);
\path[->] (Z) edge (D);
\path[->] (E) edge[bend left] (D);
\end{tikzpicture}
\end{center}
We assume that the direct path $X \ra Y$ is actually mediated by an unmeasured random variable $U$ that is left unmodelled. This arguably holds in most domains. In the example above, $U$ may represent the hirer's predisposition to hire the applicant. However, the identity of $U$ is irrelevant. Let $G$ denote the causal graph below, i.e. the original causal graph refined with the addition of $U$.
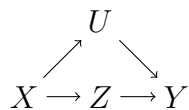
\begin{center}
\begin{tikzpicture}[inner sep=1mm]
\node at (0,0) (E) {$X$};
\node at (2,0) (D) {$Y$};
\node at (1,1) (U) {$U$};
\node at (1,0) (Z) {$Z$};
\path[->] (E) edge (Z);
\path[->] (Z) edge (D);
\path[->] (E) edge (U);
\path[->] (U) edge (D);
\end{tikzpicture}
\end{center}
Now, deactivating the direct path $X \ra Y$ in the original causal graph can be achieved by adjusting for $U$ in $G$, i.e. $\sum_u E[Y|x,u] p(u)$. Unfortunately, $U$ is unmeasured. We propose an alternative way of deactivating $X \ra Y$. Let $H$ denote the causal graph below, i.e. the result of reversing the edge $X \ra U$ in $G$.
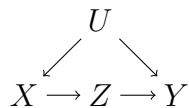
\begin{center}
\begin{tikzpicture}[inner sep=1mm]
\node at (0,0) (E) {$X$};
\node at (2,0) (D) {$Y$};
\node at (1,1) (U) {$U$};
\node at (1,0) (Z) {$Z$};
\path[->] (E) edge (Z);
\path[->] (Z) edge (D);
\path[->] (U) edge (E);
\path[->] (U) edge (D);
\end{tikzpicture}
\end{center}
The average total effect of $X$ on $Y$ in $H$ can be computed by the front-door criterion \cite{Pearl2009}:
\begin{align}\nonumber
TE(X,Y) &= E[Y_x] - E[Y_{\nx}]\\\label{eq:HTE}
&= \sum_z p(z|x) \sum_{x'} E[Y|x',z] p(x') - \sum_z p(z|\nx) \sum_{x'} E[Y|x',z] p(x').
\end{align}
Note that $G$ and $H$ are distribution equivalent, i.e. every probability distribution that is representable by $G$ is representable by $H$ and vice versa \cite{Pearl2009}. Then, evaluating the second line of the equation above in $G$ or $H$ gives the same result. If we evaluate it in $H$, then it corresponds to the part of association between $X$ and $Y$ that is attributable to the path $X \ra Z \ra Y$. If we evaluate it in $G$, then it corresponds to the part of $TE(X,Y)$ in $G$ that is attributable to the path $X \ra Z \ra Y$, because $TE(X,Y)$ in $G$ equals the association between $X$ and $Y$, since $G$ has only directed paths from $X$ to $Y$. Therefore, the second line in the equation above corresponds to the part of $TE(X,Y)$ in the original causal graph that is attributable to the path $X \ra Z \ra Y$, thereby deactivating the direct path $X \ra Y$. We propose to use the second line in the equation above as a measure of the indirect effect of $X$ on $Y$ in the original causal graph:
\[
IE(X,Y) = \sum_z p(z|x) \sum_{x'} E[Y|x',z] p(x') - \sum_z p(z|\nx) \sum_{x'} E[Y|x',z] p(x').
\]

$IE(X,Y)$ only considers the path $X \ra Z \ra Y$ to propagate the value of $X$. This is unlike $NIE(X,Y)$, which considers both paths from $X$ to $Y$: The path $X \ra Y$ propagates the value $X=\nx$, whereas the path $X \ra Z \ra Y$ propagates the value that $Z$ takes under $X=x$ and $X=\nx$. As shown in Section \ref{sec:lte}, provided that $Z$ is binary, $IE(X,Y)$ can be written as $TE(X,Z) \cdot TE(Z,Y)$, which some may find natural. It moreover coincides with the indirect effect in linear structural equation models.

Likewise, we propose to measure the direct effect of $X$ on $Y$ as the part of $TE(X,Y)$ in the original causal graph that remains after deactivating the path $X \ra Z \ra Y$. This is achieved by simply adjusting for $Z$:
\[
DE(X,Y) = \sum_{z} E[Y|x,z] p(z) - \sum_z E[Y|\nx,z] p(z).
\]
For the same reasons as above, this is unlike the measure proposed in \cite{Pearl2001,Pearl2009}, namely the average natural (or pure) direct effect $NDE(X,Y)=E[Y_{x,Z_{\nx}}]-E[Y_{\nx}]$.

Finally, note that $DE(X,Y)$ and $IE(X,Y)$ can be computed from the observed data distribution $p(X,Z,Y)$. This is also true for $NDE(X,Y)$ and $NIE(X,Y)$ \cite{Pearl2001,Pearl2009}. Note also that the sum of $DE(X,Y)$ and $IE(X,Y)$ does not equal $TE(X,Y)$ in the original causal graph, due to interactions in the outcome model. This is also true for the sum of $NDE(X,Y)$ and $NIE(X,Y)$ \cite{Pearl2001,Pearl2009,VanderWeele2013,VanderWeele2014}.

\subsection{Example}

Consider the following example from \cite{Pearl2012}, where the causal graph is
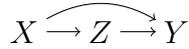
\begin{center}
\begin{tikzpicture}[inner sep=1mm]
\node at (0,0) (E) {$X$};
\node at (2,0) (D) {$Y$};
\node at (1,0) (Z) {$Z$};
\path[->] (E) edge (Z);
\path[->] (Z) edge (D);
\path[->] (E) edge[bend left] (D);
\end{tikzpicture}
\end{center}
such that $X$ represents a drug treatment, $Z$ the presence of a certain enzyme in a patient's blood, and $Y$ recovery. Moreover, we have that
\begin{align*}
p(z|x)=0.75 && p(y|x,z)=0.8\\
&& p(y|x,\nz)=0.4\\
p(z|\nx)=0.4 && p(y|\nx,z)=0.3\\
&& p(y|\nx,\nz)=0.2.
\end{align*}

\citeauthor{Pearl2012} imagines a scenario where someone proposes developing a cheaper drug that is equal to the existing one except for the lack of effect on enzyme production. To evaluate the new drug's performance, he computes $TE(X,Y)=0.46$ and $NDE(X,Y)=0.32$, and concludes that the new drug will reduce the probability of recovery by $30 \%$, i.e. $1-NDE(X,Y)/TE(X,Y)=0.3$. We can repeat the analysis using $DE(X,Y)$ instead of $NDE(X,Y)$:
\begin{align*}
DE(X,Y) &= p(y|x,z) p(z) + p(y|x,\nz) p(\nz) - p(y|\nx,z) p(z) - p(y|\nx,\nz) p(\nz)\\
&= 0.8 p(z) + 0.4 (1-p(z)) - 0.3 p(z) - 0.2 (1-p(z))\\
&= 0.2 + 0.3 p(z)\\
&= 0.2 + 0.3 [p(z|x)p(x) + p(z|\nx)p(\nx)]\\
&= 0.2 + 0.3 [0.75 p(x) + 0.4 (1-p(x))]\\
&= 0.32 + 0.11 p(x)
\end{align*}
which implies that $0.32 \leq DE(X,Y) \leq 0.43$. An interval is returned because $p(X)$ is not given in the original example (it is not needed to compute $NDE(X,Y)$ or $NIE(X,Y)$). Therefore, we conclude that the new drug will reduce the probability of recovery by between $7 \%$ and $30 \%$, depending on $p(X)$.

The new drug development scenario described above corresponds to the following alternative causal graph:
\begin{center}
\begin{tikzpicture}[inner sep=1mm]
\node at (0,0) (E) {$X$};
\node at (2,0) (D) {$Y$};
\node at (1,0) (Z) {$Z$};
\path[o->] (Z) edge (E);
\path[->] (Z) edge (D);
\path[->] (E) edge[bend left] (D);
\end{tikzpicture}
\end{center}
where the edge $X \ao Z$ means that there is an edge $X \la Z$ or $X \aa Z$. The former represents that the presence of enzyme may have an effect on the patient taking the treatment, and the latter represents the potential existence of an unmeasured confounder between them. The drug performance in this alternative causal graph is simply $TE(X,Y)$, which can be computed by adjusting for $Z$, and thus it coincides with $DE(X,Y)$ in the original causal graph, since the two graphs are distribution equivalent. In other words, it is $DE(X,Y)$ rather than $NDE(X,Y)$ that should be used to answer the original question. Note that $DE(X,Y)=NDE(X,Y)=0.32$ if and only if $p(x)=0$, i.e. everyone is untreated. This is no coincidence because $NDE(X,Y)$ in the original causal graph coincides with the average effect of the treatment among the untreated in the alternative graph \cite{OgburnandVanderWeele2012b},\footnote{\cite{OgburnandVanderWeele2012b} prove the equivalence when $X \la Z$, but the proof also applies when $X \aa Z$.} rather than with $TE(X,Y)$ which is the correct answer to the original question.

\citeauthor{Pearl2012} also imagines a scenario where someone proposes developing a cheaper drug that is equal to the existing one except for the lack of direct effect on recovery, i.e. it just stimulates enzyme production as much as the existing drug. To evaluate the new drug's performance, he computes $TE(X,Y)=0.46$ and $NIE(X,Y)=0.04$, and concludes that the new drug will reduce the probability of recovery by $91 \%$, i.e. $1-NIE(X,Y)/TE(X,Y)=0.91$.\footnote{The small disagreements with the results in \cite{Pearl2012} are due to rounding.} We can repeat the analysis using $IE(X,Y)$ instead of $NIE(X,Y)$:
\begin{align*}
IE(X,Y) &= p(z|x) [ p(y|x,z) p(x) + p(y|\nx,z) p(\nx)]\\
&+ p(\nz|x) [ p(y|x,\nz) p(x) + p(y|\nx,\nz) p(\nx)]\\
&- p(z|\nx) [ p(y|x,z) p(x) + p(y|\nx,z) p(\nx)]\\
&- p(\nz|\nx) [ p(y|x,\nz) p(x) + p(y|\nx,\nz) p(\nx)]\\
&= 0.75 [ 0.8 p(x) + 0.3 (1-p(x))] + 0.25 [0.4 p(x) + 0.2 (1-p(x))]\\
&- 0.4 [ 0.8 p(x) + 0.3 (1-p(x))] - 0.6 [0.4 p(x) + 0.2 (1-p(x))]\\
&= 0.04 + 0.11 p(x)
\end{align*}
which implies that $0.04 \leq IE(X,Y) \leq 0.15$. Therefore, we conclude that the new drug will reduce the probability of recovery by between $67 \%$ and $91 \%$, depending on $p(X)$.

The latest new drug development scenario corresponds to the following alternative causal graph:
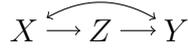
\begin{center}
\begin{tikzpicture}[inner sep=1mm]
\node at (0,0) (E) {$X$};
\node at (2,0) (D) {$Y$};
\node at (1,0) (Z) {$Z$};
\path[->] (E) edge (Z);
\path[->] (Z) edge (D);
\path[<->] (E) edge[bend left] (D);
\end{tikzpicture}
\end{center}
where the edge $X \aa Y$ represents the potential existence of an unmeasured treatment-outcome confounder. The drug performance in this alternative causal graph is simply $TE(X,Y)$, which can be computed by the front-door criterion, and thus it coincides with $IE(X,Y)$ in the original causal graph, under our assumption that the direct path $X \ra Y$ in the original graph is mediated by an unmeasured random variable (recall the previous section). In other words, it is $IE(X,Y)$ rather than $NIE(X,Y)$ that should be used to answer the original question. Note that $IE(X,Y)=NIE(X,Y)=0.04$ if and only if $p(x)=0$. Again, this is no coincidence because $NIE(X,Y)$ in the original causal graph coincides with the average effect of the treatment among the untreated in the alternative graph \cite{Pearl2001,ShpitserandPearl2009}, rather than with $TE(X,Y)$ which is the correct answer to the original question.

\subsection{Unmeasured Confounding}\label{sec:confounding}

In this section, we study the following extension of the original causal graph with an unmeasured treatment-outcome confounder $V$.
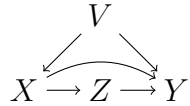
\begin{center}
\begin{tikzpicture}[inner sep=1mm]
\node at (0,0) (E) {$X$};
\node at (2,0) (D) {$Y$};
\node at (1,0) (Z) {$Z$};
\node at (1,1) (V) {$V$};
\path[->] (E) edge (Z);
\path[->] (Z) edge (D);
\path[->] (E) edge[bend left] (D);
\path[->] (V) edge (E);
\path[->] (V) edge (D);
\end{tikzpicture}
\end{center}

Now, neither $NDE(X,Y)$ nor $NIE(X,Y)$ nor their total and controlled counterparts are identifiable from the observed data distribution $p(X,Z,Y)$ \cite{Pearl2001,Pearl2009}. However, $IE(X,Y)$ can be computed pretty much like before. First, we add the unmeasured mediator $U$. The edge $V \oa U$ means that there may be an edge $V \ra U$ or $V \aa U$.
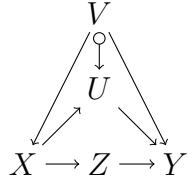
\begin{center}
\begin{tikzpicture}[inner sep=1mm]
\node at (0,0) (E) {$X$};
\node at (2,0) (D) {$Y$};
\node at (1,1) (U) {$U$};
\node at (1,0) (Z) {$Z$};
\node at (1,2) (V) {$V$};
\path[->] (E) edge (Z);
\path[->] (Z) edge (D);
\path[->] (E) edge (U);
\path[->] (U) edge (D);
\path[->] (V) edge (E);
\path[->] (V) edge (D);
\path[o->] (V) edge (U);
\end{tikzpicture}
\end{center}
Then, we group $U$ and $V$. Note that every probability distribution that is representable by the graph above is representable by the graph below, since all the independencies entailed by the latter hold in the former.
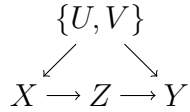
\begin{center}
\begin{tikzpicture}[inner sep=1mm]
\node at (0,0) (E) {$X$};
\node at (2,0) (D) {$Y$};
\node at (1,1) (U) {$\{U,V\}$};
\node at (1,0) (Z) {$Z$};
\path[->] (E) edge (Z);
\path[->] (Z) edge (D);
\path[->] (U) edge (E);
\path[->] (U) edge (D);
\end{tikzpicture}
\end{center}
Finally, we apply the front-door criterion.

Like $NDE(X,Y)$ and its total counterpart, $DE(X,Y)$ is not identifiable from the observed data distribution $p(X,Z,Y)$ in the extended causal graph under consideration. However, it may be bounded if $V$ is binary and a binary proxy $W$ of $V$ is measured. The causal graph under consideration is then as follows.
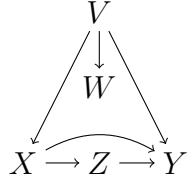
\begin{center}
\begin{tikzpicture}[inner sep=1mm]
\node at (0,0) (E) {$X$};
\node at (2,0) (D) {$Y$};
\node at (1,0) (Z) {$Z$};
\node at (1,2) (V) {$V$};
\node at (1,1) (W) {$W$};
\path[->] (E) edge (Z);
\path[->] (Z) edge (D);
\path[->] (E) edge[bend left] (D);
\path[->] (V) edge (E);
\path[->] (V) edge (D);
\path[->] (V) edge (W);
\end{tikzpicture}
\end{center}

In the literature, there are many cautionary tales about the bias that adjusting for the proxy of an unmeasured confounder introduces to the estimation of a causal effect \cite{austin2004inflation,altman2006cost,chen2007biased}. For instance, \cite{brenner1997potential} constructs an example where adjusting for the proxy is worse than not adjusting at all. However, there are conditions under which the opposite is true \cite{Gabrieletal.2022,OgburnandVanderWeele2012a,Penna2020,Sjolanderetal.2022}. We use some of these conditions here.

Recall that $DE(X,Y)$ is the part of $TE(X,Y)$ in the causal graph that remains after deactivating the path $X \ra Z \ra Y$. This is achieved by simply adjusting for $Z$:
\[
DE(X,Y) = \sum_{z} TE(X,Y|z) p(z)
\]
where $TE(X,Y|z)$ is the average total effect of $X$ on $Y$ in the stratum $Z=z$. Let us define the observed or partially adjusted average total effect of $X$ on $Y$ in the stratum $Z=z$ as
\[
TE_{obs}(X,Y|z) = \sum_{w} E[Y|x,z,w] p(w|z) - \sum_{w} E[Y|\nx,z,w] p(w|z).
\]
Note that it can be computed from the observed data distribution $p(X,W,Z,Y)$. Rephrasing the results in \cite{OgburnandVanderWeele2012a,Penna2020} to our scenario, if $E[Y|x',z',W]$ and $E[X|z',W]$ are one nonincreasing and the other nondecreasing in $W$ for all $x' \in \{x,\nx\}$ and $z' \in \{z,\nz\}$, then $TE(X,Y|z') \geq TE_{obs}(X,Y|z')$ for all $z' \in \{z,\nz\}$. On the other hand, if $E[Y|x',z',W]$ and $E[X|z',W]$ are both nonincreasing or both nondecreasing in $W$ for all $x' \in \{x,\nx\}$ and $z' \in \{z,\nz\}$, then $TE_{obs}(X,Y|z') \geq TE(X,Y|z')$ for all $z' \in \{z,\nz\}$. Note that the antecedents of these rules are testable from the observed data distribution $p(X,W,Z,Y)$. Not in all but in many cases, these rules enable us to bound $DE(X,Y)$ and even determine its sign. Specifically,
\begin{align*}
DE(X,Y) \cdot (2 \cdot \pmb{1}_{\neq} -1) \geq [ \sum_{z} TE_{obs}(X,Y|z) p(z) ] \cdot (2 \cdot \pmb{1}_{\neq} -1)
\end{align*}
where $\pmb{1}_{\neq}$ is 1 (respectively, 0) if $E[Y|x',z',W]$ and $E[X|z',W]$ are one nonincreasing and the other nondecreasing (respectively, both nonincreasing or both nondecreasing) in $W$ for all $x' \in \{x,\nx\}$ and $z' \in \{z,\nz\}$.

\comments{
\subsection{Non-Compliance} 

This section addresses a problem of randomized controlled trials, namely non-compliance with the randomly assigned treatment due to e.g. side effects. Consider the following causal graph, where $R$ denotes the randomized treatment assigned, $X$ the actual treatment received, $W$ the self-reported treatment receive, $Z$ the mediator, and $Y$ the outcome.
\begin{center}
\begin{tikzpicture}[inner sep=1mm]
\node at (0,0) (E) {$X$};
\node at (2,0) (D) {$Y$};
\node at (1,0) (Z) {$Z$};
\node at (-1,0) (R) {$R$};
\node at (0,-1) (W) {$W$};
\path[->] (E) edge (Z);
\path[->] (Z) edge (D);
\path[->] (R) edge (E);
\path[->] (E) edge (W);
\path[->] (E) edge[bend left] (D);
\end{tikzpicture}
\end{center}

Our setup above resembles the indirect experimentation setup studied by \cite{Pearl2009}, where each individual is encouraged (rather than forced) to take a randomly assigned treatment $R$ but it is the individual who ultimately selects the treatment $X$ that she receives. However, there are two main differences. First, $X$ is unmeasured in our setup. Second, our causal graph has the edge $X \ra Y$ instead of $X \aa Y$, due the focus of this manuscript being on direct and indirect effects. Specifically, we show below how to bound $IE(X,Y)$ provided that $Z$ is binary.
}

\subsection{Long-Term Effects}\label{sec:lte}

This section addresses a problem of randomized controlled trials, namely long-time effect estimation from typically short-lived trials. Consider the following causal graph, where $X$ and $V$ are unmeasured.
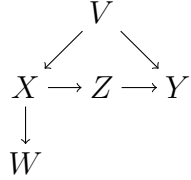
\begin{center}
\begin{tikzpicture}[inner sep=1mm]
\node at (0,0) (E) {$X$};
\node at (2,0) (D) {$Y$};
\node at (1,1) (U) {$V$};
\node at (1,0) (Z) {$Z$};
\node at (0,-1) (W) {$W$};
\path[->] (E) edge (Z);
\path[->] (Z) edge (D);
\path[->] (U) edge (E);
\path[->] (U) edge (D);
\path[->] (E) edge (W);
\end{tikzpicture}
\end{center}
We assume that the mediator $Z$ is a short-term effect of the treatment $X$, whereas $Y$ is a long-term effect of $X$. Randomized controlled trials are typically short-lived due to cost considerations and, thus, they are typically conducted to estimate short-term effects but not longer ones. Observational data, on the other hand, is much cheaper to obtain and, thus, they may include long-term outcome observations. Unfortunately, observational data is typically subject to unmeasured confounding, and mismeasurements due to self-reporting. Therefore, we assume that a randomized controlled trial was conducted to estimate $TE(X,Z)$, but not $TE(Z,Y)$ or $TE(X,Y)$. We also assume that the probability distribution $p(W,Z,Y)$ was estimated from observational data, where $W$ represents the self-reported treatment, which may differ from the actual unmeasured treatment $X$. Our goal is computing $TE(X,Y)$. Unfortunately, this cannot be done from the information available. However, the fact that $TE(X,Y)=IE(X,Y)$ implies, as we show below, that $TE(X,Y)$ can be bounded sometimes.

Our setup above is similar to the one by \cite{Atheyetal.2019} with the difference that they assume no unmeasured confounding. Our setup is also close to the one by \cite{VanGoffrieretal.2023} with the difference that they consider linear and partial linear structural equation models, while we consider non-parametric models. Moreover, unlike us, these two works assume that the true treatment is available in the observational data.

Provided that $Z$ is binary, we have that
\begin{align*}
IE(X,Y) &= p(z|x) \sum_{x'} E[Y|x',z] p(x')\\
&+ p(\nz|x) \sum_{x'} E[Y|x',\nz] p(x')\\
&- p(z|\nx) \sum_{x'} E[Y|x',z] p(x')\\
&- p(\nz|\nx) \sum_{x'} E[Y|x',\nz] p(x')\\
&= [ p(z|x) - p(z|\nx) ] [ \sum_{x'} E[Y|x',z] p(x') ]\\
&+ [ p(\nz|x) - p(\nz|\nx) ] [ \sum_{x'} E[Y|x',\nz] p(x') ]\\
&= [ p(z|x) - p(z|\nx) ] [ \sum_{x'} E[Y|x',z] p(x') ]\\
&+ [ - p(z|x) + p(z|\nx) ] [ \sum_{x'} E[Y|x',\nz] p(x') ]\\
&= [ p(z|x) - p(z|\nx) ] [ \sum_{x'} E[Y|x',z] p(x') - \sum_{x'} E[Y|x',\nz] p(x') ]\\
&= [ E[Z_x] - E[Z_{\nx}] ] [ E[Y_z] - E[Y_{\nz}] ]\\
&= TE(X,Z) \cdot TE(Z,Y).
\end{align*}

Let us define the observed or partially adjusted average total effect of $Z$ on $Y$ as
\[
TE_{obs}(Z,Y) = \sum_{w} E[Y|z,w] p(w) - \sum_{w} E[Y|\nz,w] p(w).
\]
Note that it can be computed from the observed data distribution $p(W,Z,Y)$. If $E[Y|z',W]$ and $E[Z|W]$ are one nonincreasing and the other nondecreasing in $W$ for all $z' \in \{z,\nz\}$, then $TE(Z,Y) \geq TE_{obs}(Z,Y)$ \cite{OgburnandVanderWeele2012a,Penna2020}. On the other hand, if $E[Y|z',W]$ and $E[Z|W]$ are both nonincreasing or both nondecreasing in $W$ for all $z' \in \{z,\nz\}$, then $TE_{obs}(Z,Y) \geq TE(Z,Y)$ \cite{OgburnandVanderWeele2012a,Penna2020}.\footnote{In the proofs of these results, $X$ is a parent of $V$. The results also hold when $X$ is a child of $V$, since (i) every probability distribution that is representable when $X$ is a child of $V$ is also representable when $X$ is a parent of $V$ and vice versa \cite{Pearl2009}, and (ii) $TE(Z,Y)$ and $TE_{obs}(Z,Y)$ give each the same result in both cases.} Note that the antecedents of these rules are testable from the observed data distribution $p(W,Z,Y)$. Not in all but in many cases, these rules together with the knowledge of $TE(X,Z)$ enable us to bound $IE(X,Y)$ and even determine its sign. Specifically,
\begin{align*}
&IE(X,Y) \cdot (2 \cdot \pmb{1}_{\neq} -1) \cdot (2 \cdot \pmb{1}_{\geq} -1)\\
&\geq TE(X,Z) \cdot TE_{obs}(Z,Y) \cdot (2 \cdot \pmb{1}_{\neq} -1) \cdot (2 \cdot \pmb{1}_{\geq} -1)
\end{align*}
where $\pmb{1}_{\neq}$ is 1 (respectively, 0) if $E[Y|z',W]$ and $E[Z|W]$ are one nonincreasing and the other nondecreasing (respectively, both nonincreasing or both nondecreasing) in $W$ for all $z' \in \{z,\nz\}$, and $\pmb{1}_{\geq}$ is 1 if $TE(X,Z) \geq 0$ and 0 otherwise.

Finally, note that the last equation also holds if we add the edge $X \ra Y$ to the causal graph under study. To see it, simply pre-process the graph with the three transformations at the beginning of Section \ref{sec:confounding}.

\section{Discussion}\label{sec:discussion}

We have proposed new measures of direct and indirect effects. They are based on contrasting the effects of exposure and non-exposure and they do not require selecting a reference value. This makes them unlike the existing measures in the literature and, thus, suitable in cases where the existing ones are unsuitable. The opposite is also true. The new measures assume that the direct path from expose to outcome is mediated by an unmeasured random variable. Its identity is irrelevant. This arguably holds in most domains.

When there is unmeasured treatment-outcome confounding, we have shown that the new measure of indirect effect still applies, whereas the new measure of direct effect can be bounded in some cases. This also sets them apart from the existing measures. Finally, we have shown how the new measure of indirect effect can be used to sometimes bound long-term effects by combining experimental and observational data.

As future work, we note that the bounds presented here require some random variables being binary. It would be worth studying the possibility of relaxing this requirement by making use of the results in \cite{Sjolanderetal.2022,Pennaetal.2021}. 

\bibliographystyle{unsrtnat}
\bibliography{sensitivityAnalysis}

\begin{thebibliography}{21}
\providecommand{\natexlab}[1]{#1}
\providecommand{\url}[1]{\texttt{#1}}
\expandafter\ifx\csname urlstyle\endcsname\relax
  \providecommand{\doi}[1]{doi: #1}\else
  \providecommand{\doi}{doi: \begingroup \urlstyle{rm}\Url}\fi

\bibitem[Pearl(2001)]{Pearl2001}
J.~Pearl.
\newblock {Direct and Indirect Effects}.
\newblock In \emph{Proceedings of the 17th Conference on Uncertainty in
  Artificial Intelligence}, pages 411--420, 2001.

\bibitem[Pearl(2009)]{Pearl2009}
J.~Pearl.
\newblock \emph{Causality: Models, Reasoning, and Inference}.
\newblock Cambridge University Press, 2009.

\bibitem[Geneletti(2007)]{Geneletti2007}
S.~Geneletti.
\newblock {Identifying Direct and Indirect Effects in a Non-Counterfactual
  Framework}.
\newblock \emph{Journal of the Royal Statistical Society Series B},
  69:\penalty0 199--215, 2007.

\bibitem[VanderWeele et~al.(2014)VanderWeele, Vansteelandt, and
  Robins]{VanderWeeleetal.2014}
T.~J. VanderWeele, S.~Vansteelandt, and J.~M. Robins.
\newblock {Effect Decomposition in the Presence of an Exposure-Induced
  Mediator-Outcome Confounder}.
\newblock \emph{Epidemiology}, 25:\penalty0 300--306, 2014.

\bibitem[Fulcher et~al.(2020)Fulcher, Shpitser, Marealle, and
  Tchetgen~Tchetgen]{Fulcheretal.2020}
I.~R. Fulcher, I.~Shpitser, S.~Marealle, and E.~J. Tchetgen~Tchetgen.
\newblock {Robust Inference on Population Indirect Causal Effects: The
  Generalized Front Door Criterion}.
\newblock \emph{Journal of the Royal Statistical Society Series B},
  82:\penalty0 199--214, 2020.

\bibitem[VanderWeele(2013)]{VanderWeele2013}
T.~J. VanderWeele.
\newblock {A Three-Way Decomposition of a Total Effect into Direct, Indirect,
  and Interactive Effects}.
\newblock \emph{Epidemiology}, 24:\penalty0 224--232, 2013.

\bibitem[VanderWeele(2014)]{VanderWeele2014}
T.~J. VanderWeele.
\newblock {A Unification of Mediation and Interaction: A Four-Way
  Decomposition}.
\newblock \emph{Epidemiology}, 25:\penalty0 749--761, 2014.

\bibitem[Pearl(2012)]{Pearl2012}
J.~Pearl.
\newblock {The Causal Mediation Formula - A Guide to the Assessment of Pathways
  and Mechanisms}.
\newblock \emph{Prevention Science}, 13:\penalty0 426--436, 2012.

\bibitem[Ogburn and VanderWeele(2012{\natexlab{a}})]{OgburnandVanderWeele2012b}
E.~L. Ogburn and T.~J. VanderWeele.
\newblock {Analytic Results on the Bias Due to Nondifferential
  Misclassification of a Binary Mediator}.
\newblock \emph{American Journal of Epidemiology}, 176:\penalty0 555--561,
  2012{\natexlab{a}}.

\bibitem[Shpitser and Pearl(2009)]{ShpitserandPearl2009}
I.~Shpitser and J.~Pearl.
\newblock {Effects of Treatment on the Treated: Identification and
  Generalization}.
\newblock In \emph{Proceedings of the 25th Conference on Uncertainty in
  Artificial Intelligence}, pages 514--521, 2009.

\bibitem[Austin and Brunner(2004)]{austin2004inflation}
P.~C. Austin and L.~J. Brunner.
\newblock {Inflation of the Type I Error Rate when a Continuous Confounding
  Variable is Categorized in Logistic Regression Analyses}.
\newblock \emph{Statistics in medicine}, 23:\penalty0 1159--1178, 2004.

\bibitem[Altman and Royston(2006)]{altman2006cost}
D.~G. Altman and P.~Royston.
\newblock {The Cost of Dichotomising Continuous Variables}.
\newblock \emph{BMJ}, 332:\penalty0 1080, 2006.

\bibitem[Chen et~al.(2007)Chen, Cohen, and Chen]{chen2007biased}
H.~Chen, P.~Cohen, and S.~Chen.
\newblock {Biased Odds Ratios from Dichotomization of Age}.
\newblock \emph{Statistics in medicine}, 26:\penalty0 3487--3497, 2007.

\bibitem[Brenner(1997)]{brenner1997potential}
H.~Brenner.
\newblock {A Potential Pitfall in Control of Covariates in Epidemiologic
  Studies}.
\newblock \emph{Epidemiology}, 9:\penalty0 68--71, 1997.

\bibitem[Gabriel et~al.(2022)Gabriel, Pe\~{n}a, and
  Sjölander]{Gabrieletal.2022}
E.~E. Gabriel, J.~M. Pe\~{n}a, and A.~Sjölander.
\newblock {Bias Attenuation Results for Dichotomization of a Continuous
  Confounder}.
\newblock \emph{Journal of Causal Inference}, 10:\penalty0 515--526, 2022.

\bibitem[Ogburn and VanderWeele(2012{\natexlab{b}})]{OgburnandVanderWeele2012a}
E.~L. Ogburn and T.~J. VanderWeele.
\newblock {On the Nondifferential Misclassification of a Binary Confounder}.
\newblock \emph{Epidemiology}, 23:\penalty0 433--439, 2012{\natexlab{b}}.

\bibitem[Pe\~na(2020)]{Penna2020}
J.~M. Pe\~na.
\newblock {On the Monotonicity of a Nondifferentially Mismeasured Binary
  Confounder}.
\newblock \emph{Journal of Causal Inference}, 8:\penalty0 150--163, 2020.

\bibitem[Sjölander et~al.(2022)Sjölander, Pe\~{n}a, and
  Gabriel]{Sjolanderetal.2022}
A.~Sjölander, J.~M. Pe\~{n}a, and E.~E. Gabriel.
\newblock {Bias Results for Nondifferential Mismeasurement of a Binary
  Confounder}.
\newblock \emph{Statistics \& Probability Letters}, 186:\penalty0 109474, 2022.

\bibitem[Athey et~al.(2019)Athey, Chetty, Imbens, and Kang]{Atheyetal.2019}
S.~Athey, R.~Chetty, G.~W. Imbens, and H.~Kang.
\newblock {The Surrogate Index: Combining Short-Term Proxies to Estimate
  Long-Term Treatment Effects More Rapidly and Precisely}.
\newblock Technical Report 26463, National Bureau of Economic Research, 2019.

\bibitem[Van~Goffrier et~al.(2023)Van~Goffrier, Maystre, and
  Gilligan-Lee]{VanGoffrieretal.2023}
G.~Van~Goffrier, L.~Maystre, and C.~M. Gilligan-Lee.
\newblock {Estimating Long-Term Causal Effects from Short-Term Experiments and
  Long-Term Observational Data with Unobserved Confounding}.
\newblock In \emph{Proceedings of the 2nd Conference on Causal Learning and
  Reasoning}, 2023.

\bibitem[Pe\~{n}a et~al.(2021)Pe\~{n}a, Balgi, Sjölander, and
  Gabriel]{Pennaetal.2021}
J.~M. Pe\~{n}a, S.~Balgi, A.~Sjölander, and E.~E. Gabriel.
\newblock {On the Bias of Adjusting for a Non-Differentially Mismeasured
  Discrete Confounder}.
\newblock \emph{Journal of Causal Inference}, 9:\penalty0 229--249, 2021.

\end{thebibliography}

\end{document}